\documentclass[12pt]{article} 
\usepackage{graphics} 
\usepackage{cite}
\usepackage{epsfig}
\usepackage{amssymb}
\usepackage{amsmath}

\newcommand  {\PNAS}     {{\it Proc.\ Natl.\ Acad.\ Sci.\ USA\ }}

\def\eps{\epsilon}
\def\cH{{\cal H}}

\def\cC{{\cal C}}

\begin{document}
\begin{titlepage}

\vskip 1.in
\begin{center}
\textbf{\Large Study of a model for the folding of a small protein}\\[3.em]
\textbf{\large{Andrea Nobile and Federico Rapuano}}\\[1.em]
Dipartimento di Fisica, Universit\`a di Milano-Bicocca and \\
INFN, Sezione di Milano, Italy
\end{center}
\vskip 2 cm
\begin{center}\textbf{Abstract}\end{center}
We describe the results obtained from an improved model for protein
folding.  We find that a good agreement with the native structure of a
46 residue long, five-letter protein segment is obtained by carefully
tuning the parameters of the self-avoiding energy.  In particular we
find an improved free-energy profile. We also compare the efficiency of multidimensional replica exchange method with the widely used parallel tempering.  
\end{titlepage}


\section{Introduction}

In this paper we report on the results obtained by modifying
the model proposed in \cite{Lund1}.
We show that the free-energy profile strongly depends on small changes of
the self-avoid interaction. Despite the success of the original model in 
reproducing the native structure of a small three-helix protein (10-55 fragment of the B domain of
staphylococcal Protein A), a difficulty arises in distinguishing
between two quasi-specular topologies of the native structure. 
The third helix can be either in front or 
behind the structure formed by the first and the second helix.  
The tuning of the self-avoid interaction solves this situation.
To simulate the thermodynamical properties of the system, we use a
multidimensional version of parallel tempering in which both the
temperature and others parameters of the model become dynamical variables.
This method allows for a much deeper search in the configurational space,
respect to standard tempering algorithms.

\section{The model}

In this section we shortly describe the model originally proposed in 
\cite{Lund1} which is the starting point of our study.

\paragraph{Geometrical structures.}
There are three simplified geometrical representations for the
amino acids: one for proline, one for glycine and one for all other
ammino acids.  
The configuration of the protein is determined by the two Ramachandran
angles $\phi$, $\psi$ \cite{prt}.  Bond lenghts and other bond angles are fixed. 
Thus Ramachandran angles are the only
configurational degrees of freedom of the model.
The three representations of the aminoacids have the following
charcteristics:
\begin{itemize}
\item for all aminoacids the backbone is
  represented by an $N$, $C_\alpha$ and $C'$ chain.  The $O$ and
  $H$ atoms form the hydrogen bonds and are attached to the $C'$ and
  $N$ atoms of the backbone.  The side chains are represented by a big
  $C_\beta$ atom bonded to the $C_\alpha$ atom;
\item glycine has the same representation, except
  that the $C_\beta$ atom is missing;
\item proline also has the same representation,
  except that the $H$ atom is replaced by $C_\delta$ and the
  Ramachandran angle $\phi$ is fixed. Thus the position of the
  $C_{\delta}$ atom is fixed with respect to $N$, $C_{\alpha}$ and
  $C_{\beta}$.
\end{itemize}

\paragraph{The Hamiltonian.}
The 20 different types of aminoacids are subdivided into three
hydrophobicity classes: hydrophobic (H), polar (P) and intermediate (A).
Amino acids Leu, Ile and Phe belong to class H, Ala belongs to 
class A, while Arg, Asn, Asp, Gln, Glu, His, Lys, Pro, Ser and Tyr
belong to class P. 
The Hamiltonian $\cH$ of the model is written as a sum of four terms:
\begin{equation}
\label{eq:E}
\cH=E_{\phi \psi}+E_{sa}+E_{hb}+E_{hf}\,.
\end{equation}
Here $E_{\phi \psi}$ depends only on the 
Ramachandran angles $\phi$ and $\psi$.  All other terms account for
interactions between couples of atoms. In particular
$E_{sa},\,E_{hb},\,E_{hf}$ correspond to the self avoiding, hydrogen
bond and the hydrophobic interaction. 

The term depending on the Ramachandran angles is given by
\begin{equation}
\label{eq:ER}
E_{\phi \psi}=\frac{\eps_{\phi}}{2} 
\sum _{i}(1+\cos 3\phi _{i}) + 
\frac{\eps_{\psi}}{2}\sum _{i}(1+\cos 3\psi _{i})\,,
\end{equation}
where the sum runs over all amino acids. 
All the parameters of the model can be found in \cite{Lund1}

All other terms have the general form
\begin{equation}
\label{eq:All}
\begin{split}
&E_A=\frac{\eps_A}{2}\sum_{i\ne j}
F^{ij}_A(r_{ij})\,\theta(r^c_A-r_{ij})\,,
\qquad A=\{sa,\,hb,\,hf\}\,,\\
&F_A(r)=f_A(r)-f_A(r^c_A)-(r-r^c_A)f'_A(r^c_A)\,,
\end{split}
\end{equation}
where the sum runs over all the atoms, $r_{ij}$ is the distance between atom
$i$ and $j$, $r_A^c$ is the cutoff radius, the function $F_A(r)$ is constructed in
such a way that the interaction vanishes with its derivative at the
cutoff radius.  

We now recall the form of the function $f_A(r)$ for the various
interactions.
\begin{itemize}
\item Self avoiding case ($A={sa}$).  The sum in \eqref{eq:All} in
  this case involves all atoms except the $C_\beta C_\beta$
  pair that interacts through the term $E_{hf}$
\begin{equation}
\label{eq:esa}
f_{sa}=\left(\frac{\sigma _{i}+\sigma_j + \Delta\sigma_{ij}}{r_{ij}}
\right) ^{12}\,.
\end{equation}
$\Delta\sigma _{ij}$ vanishes for all couples except for $C_\beta C'$, $C_\beta N$ and $C_\beta O$. 

\item Hydrogen bond term ($A={hb}$).  
The sum in \eqref{eq:All} runs over $ij$ where $i$ and $j$ label respectively $H$ and $O$ atoms.
The function $f_{hb}(r)$ is given by
\begin{equation} 
\label{eq:hb}
f_{hb}(r_{ij})=2u(r_{ij})v(\alpha_{ij},\beta_{ij})\,,
\end{equation}
where $\alpha_{ij}$ and $\beta_{ij}$ are the $NHO$ angles and
$HOC'$ respectively and
\begin{equation}
\begin{split}
&u(r)= 5\left(\frac{\sigma_{hb}}{r}\right)^{12} - 
6\left(\frac{\sigma_{hb}}{r}\right)^{10}\,,\\[1.em]
&v(\alpha_{ij},\beta_{ij})=\left\{
\begin{array}{ll}
\cos^{2}\alpha_{ij}\cos^{2}\beta_{ij} & 
{\textrm{if } \alpha_{ij},\beta_{ij} > \pi/2 }\\[.5 em]
0 & \textrm{elsewhere}
\end{array} \right.
\end{split}
\end{equation}

\item Hydrophobicity energy term ($A={hf}$).  The sum in \eqref{eq:All}
  is over all $C_{\beta}$ atoms belonging to the classes HH, HA, AH.
  The function $f_{hf}(r)$ is
\begin{equation}
\label{eq:hf}
f_{hf}(r)=\left(\frac{\sigma_{hf}}{r}\right)^{12} 
- 2\left(\frac{\sigma_{hf}}{r}\right)^{6}.
\end{equation}

\end{itemize}

\section{Improvement of the model}
After having reproduced all results obtained in \cite{Lund1}, we have improved the model in two respects.
First by using a more elaborate algorithm we
have explored a larger region of the parameter space in order to test
the stability of the results.

Second by modifying the self-avoid interaction we solve the problem 
of the quasi-specular degeneracy.
We discuss both items in the following

\paragraph{Computational methods.}
To simulate the model we use a multidimensional extension of parallel
tempering (multidimensional replica exchange method).  
In the standard parallel tempering \cite{stempering} \cite{partempering} various copies of
the system are simulated simultaneously with different temperatures
$\beta$ for a fixed number of steps before an exchange of the
temperatures between systems is proposed.  
In the multidimensional version \cite{mdre}, the copies of the system are evolved
with different temperatures and and with different values of the parameters. In our
study we have used as dynamical variables the temperature  and
the parameter $\eps_{hf}$ .

We denote by $\beta_1\,,\cdots \beta_m$ and $\eps_1\,,\cdots
\eps_{m'}$ the set of temperatures and of $\eps_{hf}$ considered and 
$\cC_{ij}$ the corresponding configurations weighted with the
Boltzman-Gibbs factor $e^{-\beta_i\,\cH_j(\cC_{ij})}$ (here $\cH_j$ is
the Hamiltonian with parameter $\eps_{hf}=\eps_j$). 

After a certain number of  Monte Carlo steps to be determined  by the dynamics, 
one proposes a sequence of exchanges between two pairs of parameters $\{\beta_i\,,\eps_j\}$
and $\{\beta_{i'}\,,\eps_{j'}\}$ and the two corresponding systems
$\cC_{ij}$ and $\cC_{i'j'}$ 
\begin{equation}
\begin{split}
&\{\beta_i,\eps_j,\cC_{ij}\}\to \{\beta_i,\eps_j,\cC_{i'j'}\}\,,\\
&\{\beta_{i'},\eps_{j'},\cC_{i'j'}\}\to \{\beta_{i'},\eps_{j'},\cC_{ij}\}\,,
\end{split}
\end{equation}
with probability 
\begin{equation}
\label{eq:exch}
\begin{split}
&P=\min \left(1,\,e^{-R}\right)\,,\\
&R=\beta_{i}H_{j}(\cC_{i'j'})
+\beta_{i'}H_{j'}(\cC_{ij})
-\beta_{i}H_{j}(\cC_{ij})
-\beta_{i'}H_{j'}(\cC_{i'j'})\,.
\end{split}
\end{equation} 

Each new system, which is already termalized, runs again for the same number of
Monte Carlo steps before undergoing a new exchange.

We have chosen $7$ values for the parameter $\beta$ and $6$
values for the parameter $\eps_{hf}$ thus having $42$ systems running
simultaneously. 

The temperatures are assigned by the rule
\begin{equation}
\label{eq:epsi}
T_i=T_{min}\bigg(\frac{T_{max}}{T_{min}}\bigg)^{\frac{i}{N_{T}-1}}
\end{equation}
where $N_{T}$ is the number of values.
The values of the parameter $\epsilon_{hf}$ are chosen using the rule
\begin{equation}
\label{eq:ti}
\epsilon_i=\epsilon_{min}+\frac{(\epsilon_{max}-\epsilon_{min})}{N_\epsilon-1} i
\end{equation}

We have taken the following values
\begin{equation}
\begin{split}
&\beta^{-1}=0.44,\> 0.486,\> 0.0537,\> 0.593,\> 0.655,\> 0.724,\> 0.8\,,\\
&\eps_{hf}=1.8,\> 2.02,\> 2.24,\> 2.46,\> 2.68,\> 2.9\,.
\end{split}
\end{equation}

We have implemented this algorithm on a cluster of 42 processing nodes
each one simulating 
the entire run with fixed parameter's $\{\beta_i,\eps_j\}$
while the configurations $\cC_{i'j'}$ are possibly exchanged. 
The exchanges are proposed every $\approx$ 7000 standard Monte Carlo updates; this value
allows a high number of exchanges.

The simulations consist in $250*10^6$ Monte Carlo steps for each processor and take about 
8 hours of the processors used (Athlon 2200+ at 1800 Mhz).

\paragraph{The self-avoid term.}

The smooth cutoff is changed by a simple discontinuous cutoff for the term $E_{sa}$ 
\begin{equation}
f(r)\rightarrow\widetilde{f}(r)=\left\{
\begin{array} {ll}
0 & {\textrm{if $r>r_c$}} \\
f(r) & \textrm{elsewhere}
\end{array} \right.
\end{equation}
The parameter $\Delta\sigma_{ij}$ takes the value 0.425 \AA  \ instead of 0.625 \AA.

\section{Results}

As stated in \cite{Lund1} we find that the free-energy of the model is charactherized by two minima
corresponding to the two topologies that a three helix bundle can assume.  The correct topology is the favoured one but 
the wrong one is present with a non neglegible probability. The difficulty in distinguishing between the two topologies arisesa from the 
fact that the contact patterns between helices are very similar in the two states.

The behaviour of the model with respect to the variation of the coefficient $\eps_{hf}$ that controls the strength of 
the hydrophobicity forces is very simple: weak interaction constants correspond to free-energy profiles dominated by a totally extended helix while high values of the constant correspond to disorderd collapsed states.
In the collapsed phase, changes in the configuration of the protein are very difficult to occur because of the enormous 
amount of rejected moves due to sterical collisions. This effectively reduces the  sampling efficiency.
The standard parallel tempering or simulated tempering algorithms try to solve the problem by raising and lowering the temperature, thus effectively increasing the volume of the sampled conformational space. The main drawback of these methods
is that the raise in temperature easily destroys segments of secondary structure that are difficult to recreate either 
in an uncollapsed high temperature phase or in a collapsed low temperature phase. 
These are the main reasons that led us to choose the parameter $\eps_{hf}$ as a dynamical variable.

The indicator Q of similarity with the native structure is defined
\begin{equation}
Q=\exp(-\delta^2/100\AA^2)
\end{equation}
where $\delta$ is the root mean square deviation (rmsd), and the free-energy
\begin{equation}
F(Q)=-kT\ln P(Q)
\end{equation}
where $P(Q)$ is the probability distribution of Q.

\begin{figure}
\vspace{0mm}
\begin{center}
\includegraphics[angle=-90, scale=0.4]{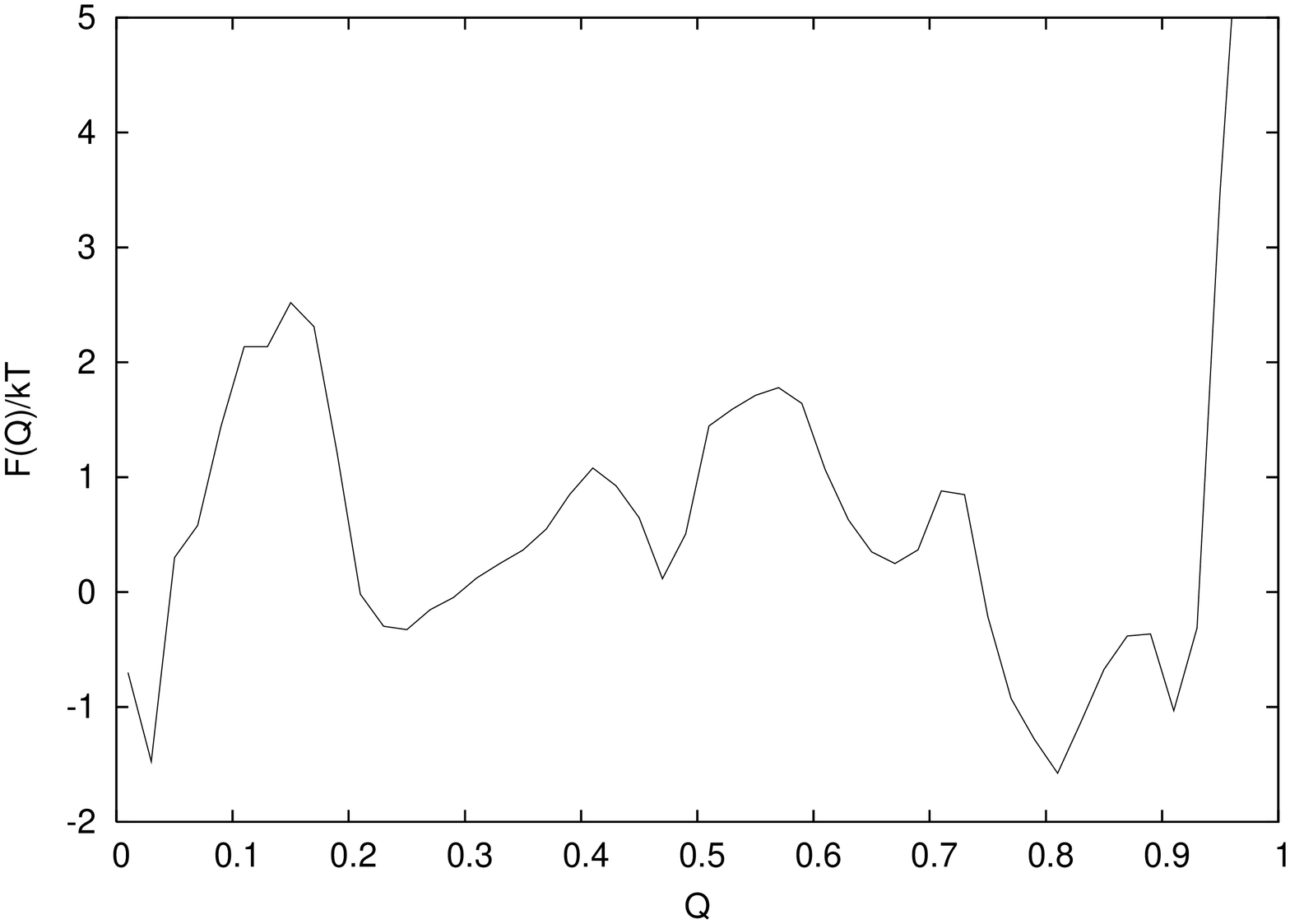}
\hspace{10mm}
\includegraphics[angle=-90, scale=0.4]{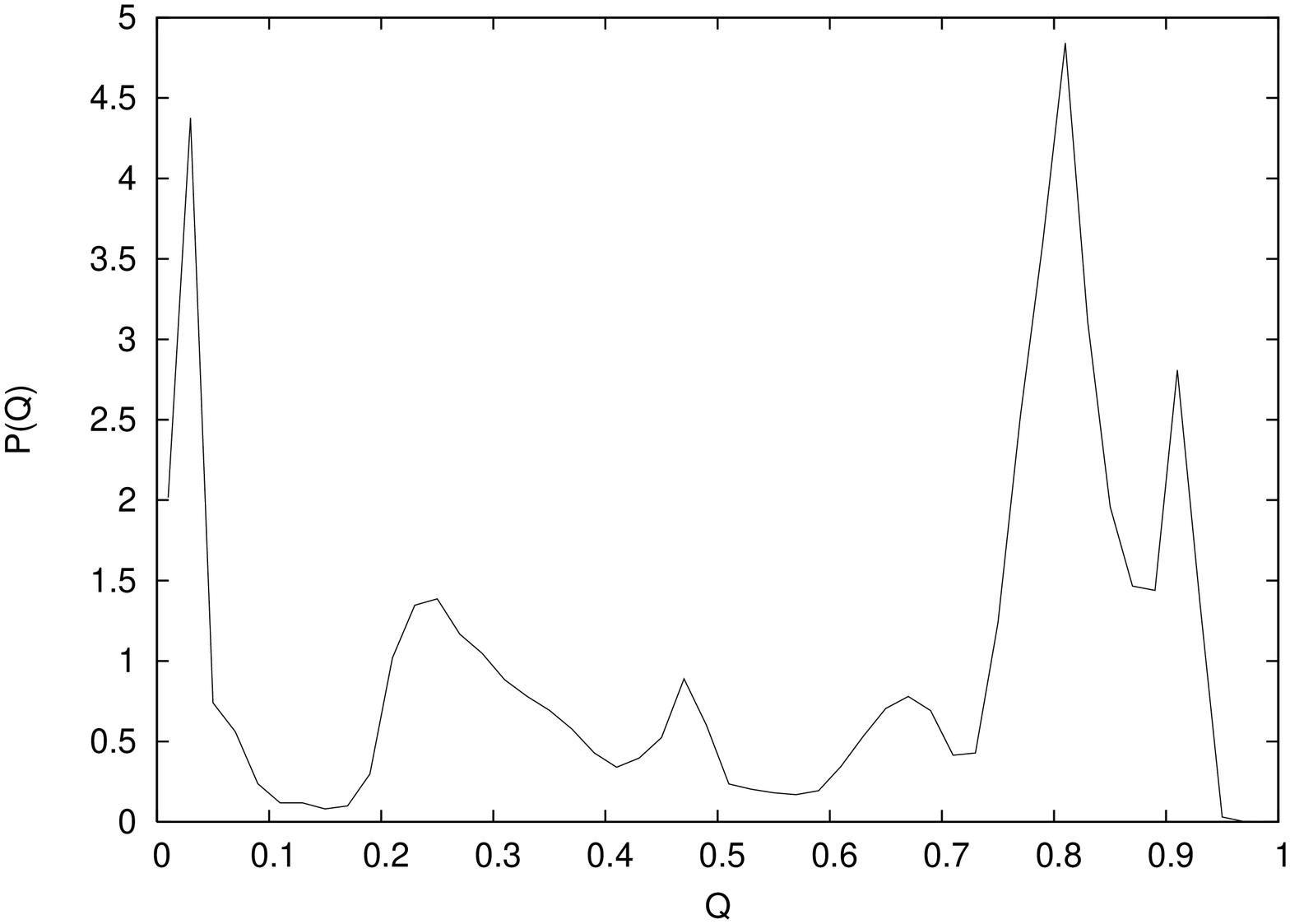}
\end{center}
\caption{\small Free-energy $F(Q)/kT$ at $kT=0.44$ and $\eps_{hf}=2.9$ (top). The distribution $P(Q)$ (bottom).
The two minima in the free-energy located at $Q\approx0.81$ and $Q\approx0.91$ correspond
to the native topology. The minima at $Q\approx0.23$ and $Q\approx0.47$ correspond to 
states composed mainly by disordered structures. The minimum at $Q\approx0.03$ is composed by structures folded in a unique long helix. }
\label{fig:2}
\end{figure}

In Fig. \ref{fig:2} we show the free energy profile $F(Q)$ and the distribution $P(Q)$ of the system at the lowest
simulated temperature $kT=0.44$ and $\eps_{hf}=2.9$.
The two most important minima at $Q\approx0.81$ and $Q\approx0.91$ correspond to the native 
topology. The structures at $Q\approx0.81$ differs from the native configuration
having the loop region between the second and the third helix partly helical and in the relative positions of the three helices. 
The first helix tends to stay more aligned with the others than in the native configuration. 
The minimum corresponding to the wrong three-helix bundle topology that was present in the original model at 
$Q\approx0.5$, is not present or neglegible in the simulations with the modified model. The minima 
at $Q\approx0.23$ and $Q\approx0.47$ correspond to disordered structures of various shapes.

\begin{figure}
\vspace{0mm}
\begin{center}
\includegraphics[angle=-90, scale=0.4]{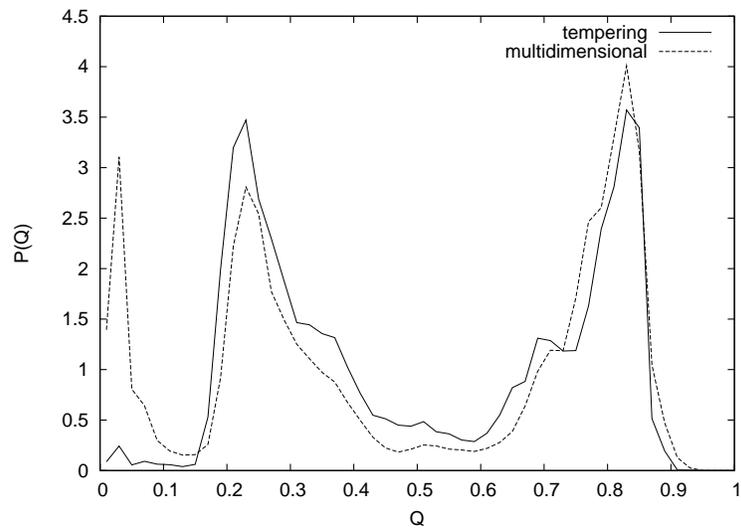}
\end{center}
\caption{\small The distribution $P(Q)$ $kT=0.44$ and $\eps_{hf}=2.9$ using parallel tempering and multidimensional replica exchange method.
The peak at $Q\approx0.03$ composed by structures folded in a unique long helix is nearly absent in the paralell tempering simulation. }
\label{fig:3}
\end{figure}

In Fig. \ref{fig:3} we compare the distribution obtained by the model using parallel tempering with
the distribution obtained using the multidiemnsional replica exchange method.
The parallel tempering simulation was done using the same two-dimensional method but setting all the allowed values for $\eps_{hf}$ to $2.9$
The model used in these simulations is characterized by a different geometry of the aminoacids
responsible for the high probability density peaked at $Q\approx0.23$. This peak corresponds to a variety of different disordered states including four helix bundles and partially disordered structures with the quasi-specular wrong topology. Our attenction foucuses on the peak at $Q\approx0.03$ wich is present in the simulation with the multidimensional replica exchange method but nearly
absent in the other. 

A reasonable explanation is that, when simulating at constant $\eps_{hf}$ , the probability 
of getting a totally helical configuration is neglegible because of the most entropically favoured 
collapsed conformations, thus leaving that region of the phase space unexplored. 
In the multidimensional replica exchange simulation, totally helical states are produced at low 
$\eps_{hf}$ values and can reach high $\eps_{hf}$ values through paths chracterized by low temperatures.
Promoting $\eps_{hf}$ to the role of dynamical variable allows for a deeper search in phase space.

Trying to suppress the totally helical state by increasing the value of $\eps_{hf}$ results in 
free-energy profiles populated by disordered and collapsed states with poor or not well-defined secondary structure.

\begin{figure}
\vspace{0mm}
\begin{center}
\includegraphics[angle=-90, scale=0.4]{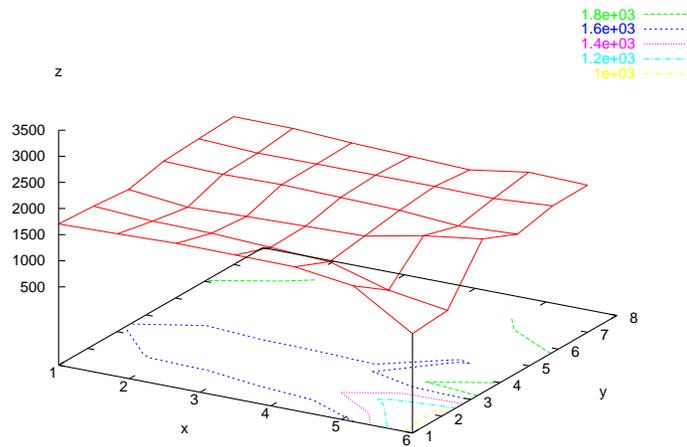}
\end{center}
\caption{\small The permanence histogram of a structure obtained using multidimensional replica exchange method.
The number of samples ($z$ axis) is plot against the index of temperature ($y$ axis) and of $\eps_{hf}$ ($x$ axis).}
\label{fig:4}
\end{figure}

In Fig. \ref{fig:4} we show a sample of a \emph{permanence histogram} relative to a structure started from an hamiltonian at hight temeprature and high $\eps_{hf}$ obtained with the multimensional replica exchange method.
The histogram is nearly flat and ensures that the algorithm is working properly.
This means that the structure walked through all the hamilonians with flat probability.

\section{Conclusions}

We have explored the properties of a modified version of the model presented in \cite{Lund1} with 
the use of the multidimensional replica exchange method. We have shown that a new self-avoiding term in the hamiltonian and a different sampling of the configurational space lead to a better agreement with the native structure.
This indicates that the model is sensitive to small changes in geometry and thus the self-avoid interaction plays a very important role 
in determining not only the local conformation (the secondary structure), but has also strong influences on the tertiary structure.

The minimum in the free-enegy profile corresponding to the totally extended helical state is an indication that the effect of the solvent 
and of the electrostatic interaction lack of detail. This result also validates the multimensional replica exchange method as a very efficient tool 
for the exploration of rugged energy landscapes.

\section{Aknowledgements}
We thank Anders Irb\"ack, Guido Tiana, Giuseppe Marchesini and Claudio Destri for fruitful 
discussions.

\end{document}